# Displacive-type ferroelectricity from  magnetic correlations within spin-chain

Tathamay Basu,[1] V. V. Ravi Kishore,[2]  Smita Gohil,[1] Kiran Singh,[1,3] N. Mohapatra,[4]  S. Bhattacharjee,[1] Babu Gonde[1], N. P. Lalla,[3] Priya Mahadevan,[2]   Shankar Ghosh,[1]  and E.V. Sampathkumaran[1]

[1]Tata Institute of Fundamental Research, Homi Bhabha Road, Colaba, Mumbai-400005, India
[2]S. N. Bose National Centre for Basic Sciences, Sector-III, Block-JD, Salt Lake, Kolkata - 700 098, India
[3]UGC-DAE Consortium for Scientific Research, University Campus, Khandwa Road, Indore - 452001, India
[4]School of Basic Sciences, Indian Institute of Technology Bhubaneshwar, Bhubaneshwar-751013, India

Correspondence and requests for materials should be addressed to E.V.S (sampath@mailhost.tifr.res.in)

**Observation of ferroelectricity among non-$d^0$ systems, which was believed for a long time an unrealistic concept, led to  various proposals for the mechanisms to  explain the same (i.e. magnetically induced ferroelectricity) during last decade. Here, we provide support for ferroelectricity of a displacive-type possibly involving magnetic ions due to short-range magnetic correlations within a spin-chain, through the demonstration of magnetoelectric coupling in a Haldane spin-chain compound $Er_2BaNiO_5$ well above its Néel temperature of ($T_N$=) 32K. There is a distinct evidence for electric polarization setting in near  60 K around which there is an evidence for short-range magnetic correlations from other experimental methods. Raman studies also establish a softening of phonon modes in the same temperature (T) range and T-dependent x-ray diffraction (XRD) patterns also reveal lattice parameters anomalies. Density-functional theory based calculations establish a displacive component (similar to $d^0$-ness) as the root-cause  of ferroelectricity from (magnetic) $NiO_6$ chain, thereby offering a new route to search for similar materials near room temperature to enable applications.**

The field of magnetoelectric (*ME*) coupling and multiferroicity involving two correlated phenomena, namely, ferroelectricity  and magnetic ordering, bearing relevance to  applications, has expanded recently. However, this kind of multiferroicity  has been considered rare in transition-metal systems, as  magnetism involves partially filled *d* orbital and ferroelectricity requires empty *d* orbital. The observation of magnetism-induced ferroelectricity  in  some  materials  led  to  the proposals of several mechanisms [1-8], for example, lone-pair stereochemistry, role of spin-spiral structure, charge ordering, and geometrical considerations. However, a displacive-type mechanism involving off-centering of the magnetic ion proposed in  Ba-doped $SrMnO_3$ [9, 10] is uncommon among stoichiometric (undoped) compounds. Here, we demonstrate that the one-dimensional short-range magnetic correlation plays an important role to induce ferroelectricity above long-range magnetic ordering in an undoped spin-chain compound possibly by this mechanism.

The compound under discussion is a Haldane spin-chain compound, $Er_2BaNiO_5$, crystallizing in an orthorhombic structure (*Immm*). We show that this family, $R_2BaNiO_5$ (R= Rare-earth), is characterized by a strong *ME* coupling and potentially prone for multiferroicity due to a displacive-type mechanism. Though this family of insulating oxides aroused considerable interest due to Haldane spin-chain magnetic anomalies [11 - 22], very little work has been reported [17, 21, 22] to understand dielectric or multiferroic behavior. Experimental studies on a few members of this family e.g. Ho [17] and Dy [22] were reported which revealed ferroelectric behaviour, but the microscopic mechanism of the observed behaviour has not been clear. Keeping this in mind, we have explored *ME* coupling on the Er member through macroscopic and microscopic measurements, augmented by Raman scattering and XRD studies, and electronic structure calculations. The orientation of the magnetic moments has been known to be decided by the anisotropy of the rare-earth ions and therefore the primary motivation with which this study was undertaken was to explore whether this factor anyway controls appearance of multiferroicity in this series. For instance, in $Dy_2BaNiO_5$, in which long-range antiferromagnetic order sets in at ($T_N$=) 58 K, the magnetic moments of Dy and Ni at 1.8 K lie almost along *c*-axis, with these moment-directions getting gradually decoupled with increasing temperature rotating towards *a*-axis (along Ni-O-Ni chain axis). On the other hand, in contrast to other R members, in the Er case, the directions of Er and Ni moments point along *a*-axis at all temperatures below $T_N$ [13], thereby suiting to the motivation of this work.

**Results**

The *T*-dependence of dielectric constant (ε') and loss factor (*tan δ*) in the presence of various magnetic fields (*H*) are shown in figure 1(a,b) for a frequency (ν) of 50 kHz . The behavior of heat-capacity (*C*) is also shown in figure 1, primarily to understand the features in complex dielectric data. In the *C/T* curve, in the absence of a magnetic field, there is a weak jump at 32 K at the onset of magnetic ordering (see inset of Fig. 1c), followed by a gradual fall with decreasing temperature till about 15 K, below which there is another feature manifesting as an upturn and a peak at 8 K. It is clear from the figure 1c that the temperatures at which the upturn and the peak occur (below 20 K) get lowered gradually with increasing magnetic-field (*H)*. A broad hump is also observed spreading over a wide *T*-range around 40 to 60 K, which could be attributable to (intrachain) short-range magnetic order, also considering that the presence of magnetic short-range order above $T_N$ was reported by a spectroscopic study in this compound [23]. Keeping these observations in mind, we turn to dielectric data. The values of *tanδ* are very low consistent with the insulating behavior of this compound. We noted a fall in ε' below 8 K for 5 kHz with a concomitant anomaly (a peak) in *tan δ* and this feature was found to exhibit a ν-dependence with this fall appearing at ~10K for 70kHz (not shown here). Similar ν-dependence was observed in *ac* susceptibility as well (not shown here) and we will not focus this aspect as this is not the aim of the article. Though it is not easy to infer a change at $T_N$ from the plots of ε'(*T*) (see, for instance, the curve shown in Fig. 1a for ν= 50 kHz) obtained in the absence of a magnetic field, a careful comparison with the plots taken in the presence of *H* reveals an anomaly around this temperature. In ε'(*T*), the curves for different *H* values essentially overlap above ~40 K, and these tend to bifurcate clearly below a temperature close to 32K, thereby offering a support for *ME* coupling. The corresponding feature observed in figure 1b (for tan*δ* (*T*)) is more intriguing, as it reveals a distinct separation of the curves measured under different *H* at a temperature much higher than $T_N$, that is, below ~ 45 K, though the observed change in the value is very small near 45 K. We attribute this to a magnetic precursor effect before the long range magnetic ordering sets in, that is, to the existence of short-range correlations proposed above. The value of ε' tends to increase with *H* below $T_N$. It is worth noting that the temperature at which a sudden drop (for the 10K-feature in figure 1a) occurs decreases with increasing *H*, for instance from 10 K for *H*= 0 to about 8 and 4 K for 50 and 140 kOe respectively for 50 kHz, with this temperature closely tracking the trend observed for the corresponding feature in the *C(T)* curves (Fig. 1c). The decrease of anomaly/peak position towards low temperature with increasing magnetic field indicates dominant antiferromagnetic component around this temperature. The observed correlation near 10K between dielectric and heat capacity results provides an additional support for the proposal of *ME* coupling in this compound. This dielectric behavior with the application of magnetic field is the same for other frequencies (5-100 kHz) as well, though the absolute values are different (not shown here).

We have also obtained further evidence for *ME* coupling effect on the basis of isothermal magnetization (*M*) behavior and ε' as a function of *H* at different temperatures, the rate of variation of *H* being 70 Oe/s (see figure 2). At 2 K, we see a distinct evidence for two jumps in *M(H)*, a stronger one near 20 kOe and the other weak one near 30 kOe (Fig. 2a). Such metamagnetic transitions were reported at 4 K as well by Chepurko et al [23]. To bring out *ME* coupling, we show Δε' as a function of *H* at 2 K in figure 2b, where Δε'= (ε'$_{H=0}$- ε'$_H$)/ε'$_{H=0}$. We find that there is a sudden upturn in the value around 20 kOe with another weak shoulder around 30 kOe, followed by a gradual non-linear rise at higher fields, similar to *M(H)*. This correlation between *M(H)* and Δε'(*H*) data establishes *ME* coupling in this compound.

The temperature dependent remnant polarization *(P)* was measured during warming in the absence of electric-field after cooling the sample in an application of electric-field (*E*) of 220 kV/m under zero magnetic field and also in the presence of different *H* for the field-cooled condition (see the method section for details). The results obtained are shown in figure 3a. Interestingly enough, though long range magnetic order sets in at 32 K, remnant polarization sets in at a much higher temperature (~ 60 K) compared to $T_N$. The sign of polarization was also found to get reversed with a reversal in the sign of electric field (inset of figure 3a), which is sufficient to claim the ferroelectric behaviour of the sample and similar ferroelectric behaviour has been reported for $Ho_2BaNiO_5$[17] and other materials also[1,2]. We endorse a possible explanation in terms of short-range low-dimensional magnetic interactions well above $T_N$ [17], which in turn facilitates ferroelectric behaviour through magnetoelectric coupling. It should be stressed that such reports bringing out multiferroic behavior due to isolated spin-chains are generally rare in the literature and we have seen such effects in another spin-chain compound $Ca_3Co_2O_6$ recently [24]. As an evidence for *ME* coupling, the values of *P* are found to change with the application of magnetic-fields well below 60 K, in addition to a very small change in the ferroelectric transition temperature (around 60 K). The plot (see Figure 3b) of pyroelectric current, $I_{pyro}$, also exhibits a peak around 55 K with its intensity and the peak-temperature varying with *H* marginally. It is worth noting that the overall variation of *P* in the magnetically ordered state is much larger than that observed for the Dy analogue [22]; we observed a value of about 4μC/m$^2$ for the Dy case by an application of 400 kV/m, whereas for the Er member, it is ~100μC/m$^2$ for an application of 220kV/m only; for the Ho case, the results were not reported [17] for the absence of a magnetic-field to enable us to compare. It is to be noted that the *P* value almost saturates below magnetic ordering $T_N$.

It is worth stating that the *dc* resistivity of this compound at room temperature is of the order of giga-ohm-cm and increases exponentially with decreasing temperature; below 120 K, it increases drastically further (inset of figure 3b) and, below 90 K, it was not possible to measure with proper resolution due to the limitations of the electrometer. This clearly establishes that this compound is a highly insulating system and whatever *ME* coupling we report here is not due to leakage current.

We have also measured the complex impedance as a function of frequency at different temperatures applying different bias voltages (0.1-2 V) using the same LCR meter and the experimental set up which was used for dielectric measurements. It was noted that the current-voltage phase angle is ~ -90° (above -89.5° below around 70 K within the measured frequency range) for a wide frequency range (between 2-500 kHz). This means that the compound is undoubtedly a very good capacitor in this range, which is a prerequisite for the investigations for the aim of this article. There is no change of feature or value with different bias voltages below 100

K. Therefore, we are confident that our conclusions from dielectric measurements are intrinsic, without any dominant influence of extrinsic effects (like contact problem, grain boundary etc.).

We have also obtained spectroscopic evidence for mode-softening attributable to the formation of polar state through Raman scattering experiments. $Er_2BaNiO_5$ forms in the space-group *Immm* at room temperature [12]. The compound has 9 atoms per unit cell and, out of the 24 vibrational modes, only the Er and the O atoms participate in the Raman modes, while all others participate in the infra-active (*IR*) modes only [25]. Figure 4a shows vibrational spectra for representative temperatures. Certain modes marked by shaded bands ($B_1, B_2, B_3$) were found to lose intensity at lower temperatures, while others persist. The persisting modes that are marked as $S_3, S_7, S_8$ are assigned the symmetry of $B_{1g}, B_{3g}$ and $A_g$ respectively and are Raman active. The modes $S_5$ and $S_7$ appear to be combinational Raman scattering modes. The mode assignment was done by comparing the present spectra with those for other compounds of the Haldane chain family [25-27]. We assign the spectral features in the shaded bands to the *IR* modes which arise from any kind of disorder in the Ni-O chain. The key observation is that there is a dramatic loss of intensity of these disorder-based modes with decreasing temperature. We have to necessarily attribute the loss of intensity of the IR modes to the gradual development of magnetic order within this chain (as inferred from the *C(T)* as well). The intensity of these magnetic-disorder induced modes is thus proportional to the mechanical susceptibility of the Ni-O chain. If the susceptibility is higher, it is more probable for thermal fluctuations to generate disorder [28]. The inverse of the intensity is thus related to the stiffness of the Ni-O chain. Figure 4b shows the inverse of the intensity of the disorder-induced modes as a function of temperature. It is apparent that there is a distinct inverted cusp around 45 K, as though there is an order-disorder phase transition of second-order type. The signature of this transition can also be seen in the variation of the peak position ($\omega_i$) of some representative peaks $S_1 ... S_9$ (marked in Fig. 4a) as a function of temperature (see Fig. 4c). The value of $\partial \omega_i / \partial T$ clearly shows a change at 60K and in the temperature range $45 < T < 60K$ (shaded region), it is positive. We associate this mode stiffening to the distortion of $NiO_6$ octahedra. Below 40K, the sign of the slope is negative. This continuous softening of the modes at low temperatures is related to the renormalization of the phonon frequencies due to the presence of strong spin-phonon coupling in this system.

The results presented above provide distinct evidence for softening of the modes in the *T*-region where short-range magnetic-ordering-induced ferroelectricity tends to set in. The observation of ferroelectricity in this family is indeed puzzling, especially since the experimentally determined crystal structure as shown in Fig. 5(a) is centro-symmetric. The Ba and rare earth ions occupy high symmetry positions in the tetragonal unit cell. In order to throw more light on this aspect, electronic structure calculations as well as phonon calculations were performed (for calculation details see in method section). The generalised gradient approximation [29] was used for the exchange-correlation functional, though we also performed calculations including an onsite Hubbard U-like term on Ni to treat electron-electron interactions in a mean-field manner. Various magnetic configurations were considered; however, the lowest energy was found for the one with antiferromagnetic one-dimensional network of Ni-O chains running along the a-direction. The coupling between chains was found to be extremely weak and ~ 0.4 meV. As a starting point, the phonon dispersions were calculated for the reported structure [13] along various symmetry directions using the small displacement method. Soft phonon modes were explored and displacements were made along the eigenvectors corresponding to the soft modes and the internal positions were again optimised. In every case, we worked with the experimental volume, optimising only internal coordinates. A soft mode was identified to develop at Γ point (Fig. 5(b)). Using the displacements suggested by the soft mode, we arrive at a lower energy structure which is non-centrosymmetric and corresponds to the space group *Imm2*. Each Ni atom is surrounded by six oxygens, though the connectivity of these $NiO_6$ motifs is present only in the a-direction through corner-shared oxygens. The primary displacement in the ferroelectric structure involves the Ni atoms moving towards a pair of oxygens in the bc-plane. As a result, (among the four equal Ni-O bond lengths), a pair of Ni-O bond-lengths become shorter while the other pair are longer (. These become 2.177 and 2.187 Å (Fig. 5(c)) when the experimental antiferromagnetic structure is imposed, while these are equal to 2.142 and 2.228 Å (Fig. 5(d)) when a ferromagnetic structure is considered. The Ni-O bond-lengths along the chain direction, remain unchanged. There are small dipole moments associated with the Ba-O and Er-O network also. The polarization is computed to be 0.39 μC/cm$^2$ for the ferromagnetic solution, while it is scaled down by almost an order of magnitude for the experimentally observed antiferromagnetic structure. Further, increasing the inter-chain separation artificially in our calculations, thereby making the chains more one-dimensional, we find an increase in the observed polarization. It is to be noted that polarization is absent if we do not introduce the one-dimensional magnetic interaction in our calculation.

In order to understand the mechanism for the observed polarization, we have examined the density of states. The system is found to be an insulator. Broadly, for any transition-metal atom in an octahedral environment, there is a lifting of degeneracy of the d-orbitals with the orbitals with $t_{2g}$ symmetry at lower energies compared to the orbitals with $e_g$ symmetry. The reduced symmetry of the NiO$_6$ octahedron in the present case ($D_{2h}$), lifts the degeneracy further; however, we can still identify the d-orbitals which have $t_{2g}$ symmetry and those which have $e_g$ symmetry and so we continue the discussion in terms of the $t_{2g}$ and $e_g$ states. The basic energy level diagram can be understood as consisting of completely filled $t_{2g}$ orbitals on Ni, and half-filled $e_g$ orbitals. As the bandwidth is smaller than the exchange splitting, the system is insulating.

The calculations described above reveal that there are changes in the Ni-O bond distances. In order to gain knowledge on the influence of ferroelectricity on the lattice constants, we have carried out x-ray diffraction studies as a function of $T$ down to 10 K. In the powder diffraction patterns, we have not observed any additional reflections in the whole temperature range, other than those known for this compound. Rietveld refinement (see supplementary information) was carried out to capture any signature of the phase transition. Refined unit-cell parameters and volume are plotted in figure 6. It is evident from this figure that the cell-parameter '*b*' indeed shows a clear upward deviation from the thermal contraction behavior with decreasing $T$ in the vicinity of ferroelectric transition temperature (around 70K), apart from the one close to $T_N$. This deviation is well-outside the error bar [~0.0001 Å]. The unit-cell parameters '*a*' and '*c*' also reveal a weak change of slope in the same temperature regions. Naturally, the unit-cell volume also shows a distinct anomaly. We have also explored whether there is any change in the lattice symmetry following ferroelectric transition by Rietveld fitting of XRD patterns for two ferroelectric subgroups, *Imm2 and I222*, of (room temperature) *Immm* space group. For *Imm2*, oxygen at 8l site [of *Immm*] splits into two 4d [of *Imm2*] Wyckoff sites, resulting in increase in refinable position parameters. For *I222*, all three position coordinates [8k site] of oxygen atom can be varied during refinement. We found that the pattern is equally well-fitted with all the three space groups (see supplementary information). Due to possible very small changes in position coordinates across the phase transition, the present Rietveld refinements can not conclusively determine the true symmetry of ferroelectric phase. Such difficulties are not unusual when the lattice distortions are weak, as remarked in Ref. 30 (also see, Ref. 7 in Ref. 30). In any case, the present XRD results provide evidence for a displacive-type ferroelectricity as indicated by changes in cell-parameters across the transition.

**Discussions**

It is important to note that the distortion observed in our calculations is similar to what has been seen in d$^0$ ferroelectrics, but rarely seen in systems with a finite d-electron count. From the details provided above, it is clear that this is not the same as the exchange-striction driven mechanism observed for Ca$_3$CoMnO$_6$ [31]. The natural question at this point is why the displacive ferroelectricity proposed here is rare. In the past, systems exhibiting non-symmetric degeneracy lifting first-order Jahn-Teller distortions [32-34] as well as band-insulators exhibiting second-order Jahn-Teller distortions [7,8, 35] have been believed to be possible candidates for displacive ferroelectricity among finite d-electron systems. The electronic structure discussed earlier puts the system into a class of band-insulators. However, every band insulator system does not turn ferroelectric and the question arises why is that the present family is so susceptible to ferroelectricity. In the case of the classic example of $BaTiO_3$ ferroelectric [36], it is known that the significant energy-gain for the ferroelectric structure comes from the band-energy gain arising from shorter Ti-O bonds [37]. For $Ni^{2+}$ oxides, one finds from an earlier work [38] by some of us that the charge-transfer energy decreased as one moves from three-dimensional NiO to a two-dimensional network of $NiO_6$ motifs and then to a one-dimensional one. A smaller charge transfer energy implies a larger band energy gain, and hence ferroelectricity seems to be likely in these one-dimensional oxides.

In short, we establish that the Haldane spin-chain compound, $Er_2BaNiO_5$, which has been previously shown to undergo long-range antiferromagnetic order below 32 K, is characterized by magnetoelectric coupling with ferrolectricity setting in above $T_N$. The key conclusion is that this family of magnetic materials is prone for multiferriocity under favorable circumstances due to the 'Displacive-type mechanism', that distorts the $O_6$ octahedra, thereby lifting the point of inversion symmetry. This is not so-commonly encountered in undoped materials with non-$3d^0$ electronic configuration. Interestingly, short-range magnetic correlations appear to be adequate to trigger ferroelectricity due to this mechanism. We hope this work triggers further work to search for similar materials near room temperature.

**Methods:**

Polycrystalline form of the compound, $Er_2BaNiO_5$, was prepared as described earlier [15] starting with stoichiometric amounts of high-purity (>99.95%) oxides, $BaCO_3$, $Er_2O_3$ and NiO. X-ray diffraction study established single phase nature of the materials.

The *dc* $\chi$ measurements were carried out using commercial Superconducting Quantum Interference Device (SQUID, Quantum Design, USA) and isothermal magnetization behavior was also performed with the help of a commercial Vibrating Sample Magnetometer (VSM, Quantum Design, USA) at various temperatures. The features in *M(T)* and *M(H)* curve are in broad agreement with those reported in the literature [13, 14, 16], thereby ensuring the quality of the sample.

Heat-capacity measurements were done by a relaxation method using commercial Physical Properties Measurement System (PPMS, Quantum Design).

Complex dielectric permittivity measurement were carried out using Agilent E4980A LCR meter with a homemade sample holder coupled to the PPMS. Temperature dependent complex dielectric permittivity was measured for various frequencies (5-100 kHz) at 1V ac bias during warming (0.5 K/min). Isothermal magnetic-field dependence of dielectric behavior was measured at different temperatures (below and above $T_N$) at different frequencies (10-100 kHz). The rate of change of magnetic-field in all these measurements (magnetization and dielectric both) is 70 Oe/sec. Remnant polarization was measured with Keithley 6517A electrometer (using the same set up like dielectric measurement integrated to PPMS) in Coulombic mode (with automated integration of $I_{pyro}$) as a function of temperature. An electric-field of 220 kV/m was applied at 80 K (above ferroelectric temperature) to align the electric dipoles and then the sample was cooled to 6 K; subsequently, the electric-field was switched off and the capacitor (sample) was shorted for sufficient time to completely remove the stray charges (if any). *P(T)* curve was then obtained during warming (2 K/min). We have also obtained these curves under different magnetic fields with the application of both *H* and *E* at 80 K and then the specimen was cooled down to the lowest measured temperature in the presence of both the fields; then polarization was measured in the same manner as discussed above, only keeping the magnetic field switched on till the end of the measurements.

The *dc* resistivity was also measured in the two-probe method using Keithley 6517A electrometer.

Raman scattering measurements were performed in the backscattering geometry using a triple grating Raman spectrometer (T64000: Jobin Yvon) equipped with a liquid nitrogen cooled charge-coupled device. The excitation source was the 647.1 nm line of a mixed gas laser (Stabilite 2018: Spectra Physics).

Low temperature Raman experiments were carried out using a continuous flow helium cryostat (Microstat He: Oxford Instruments).

Electronic structure calculations were performed within a plane-wave implementation of density functional theory within the Vienna Ab initio Simulation Package (VASP) code [39, 40]. A k-points mesh of 4x4x4 were used and an energy cutoff of 500 eV was used for determining the maximum kinetic energy of the plane waves included in the basis. The polarization of the ferroelectric structure was computed using the Berry phase method as implemented in VASP.

The x-ray diffraction patterns were taken at different temperatures in the range 10 - 300 K in a symmetric Bragg-Brentano geometry, with horizontally scanning 2θ-arm using Rigaku-make rotating anode x-ray generator operated at 12 kW. Lowering of temperature was achieved using liquid helium cryostat (Oxford Instruments, UK), which is directly mounted on a goniometer from Huber, Germany. The high quality of the data can be judged from the low and flat background with high peak to back ground ratio (which is more than 500). Rietveld refinement has been carried out using FullProf Suite package [41].

**Acknowledgements**

The authors thank Mr. S. Chowki for participation in sample preparation and Mr. Kartik K Iyer for his help in all these studies.

**Author contributions**
T.B. has carried out the magnetization, heat capacity, complex dielectric, resistivity and polarization experiments and analysed the data. B.G. and K.S. helped in polarization experiment and analysis. K.S. and N.P.L. carried out *T*-dependent XRD studies and S.B. carried out analysis of these XRD data. V.V.R.K. and P.M. has done the theoretical studies. S. Gohil and S. Ghosh carried out the Raman studies and analysed the data. N.M. and T.B. prepared and characterized the sample. E.V.S. has supervised the whole project, written the draft and finalized in consultation with other authors, in particular after discussions with P.M., S.G. (2), T.B., and K.S.

**Competing financial interests**

The authors declare no competing financial interests.

**Figure 1:** Dielectric constant and the loss factor as a function of temperature measured with a frequency of 50 kHz in the presence of external magnetic fields are plotted in (**a**) and (**b**) for $Er_2BaNiO_5$. The plots of *tanδ* shown in (b) are to be used primarily to highlight the feature near 45 K in the presence of *H*. The heat-capacity data (*C/T vs T*) measured under different magnetic fields are plotted in (c); in inset of (c), the data is plotted in a wider temperature range to highlight the weak feature around $T_N$ (marked by an arrow) and a broad anomaly above $T_N$.

**Figure 2:** Isothermal magnetization (*M(H)*) and change (Δε') in dielectric constant, $\Delta\varepsilon' = (\varepsilon'_{H=0} - \varepsilon'_H)/\varepsilon'_{H=0}$ for $Er_2BaNiO_5$.

**Figure 3:** For $Er_2BaNiO_5$, (**a**) electric polarization, *P,* as a function of temperature (in the absence and in the presence of external magnetic fields), obtained as described in the text; inset: *P* for a change in the sign of electric-field for *H*=0. (**b**) Pyroelectric current data; inset shows electrical resistivity data.

**Figure 4:** (a) Vibrational spectra at representative temperatures for $Er_2BaNiO_5$. The features corresponding to vibrational modes are shown with shaded bands. The disorder-induced infrared modes are within the shaded-band. The modes $S_3$, $S_6$, $S_7$, $S_8$ and $S_9$ are Raman modes of single scattering events, while $S_5$ and $S_7$ are Raman modes that involve multiple modes; (b) the inverse of the integrated intensity of the modes within the shaded-band as a function of temperature. The dash-dotted line is a guide to the eyes. (c) Variation of peak position of the representative modes $S_1$-$S_9$ as a function of temperature. The shaded region marks the temperature range from 60K to 45K where there is a distinct change in slope.

**Figure 5:** (a) The projection of crystal structure of $Er_2BaNiO_5$ along a-direction; (b) the phonon dispersions are shown for the symmetry directions Γ-X and Γ-L; and (c) and (d) show the Ni-O bond lengths in the ferroelectric states for antiferromagnetic and ferromagnetic structures respectively, as obtained in the calculations.

**Figure 6:** The lattice constants, *a, b*, and *c*, as well as unit-cell volume (*V*) for $Er_2BaNiO_5$ as a function of temperature determined from x-ray diffraction pattern. The vertical dashed-lines are drawn to show the *T*-region around which lattice constants reveal some anomalies.

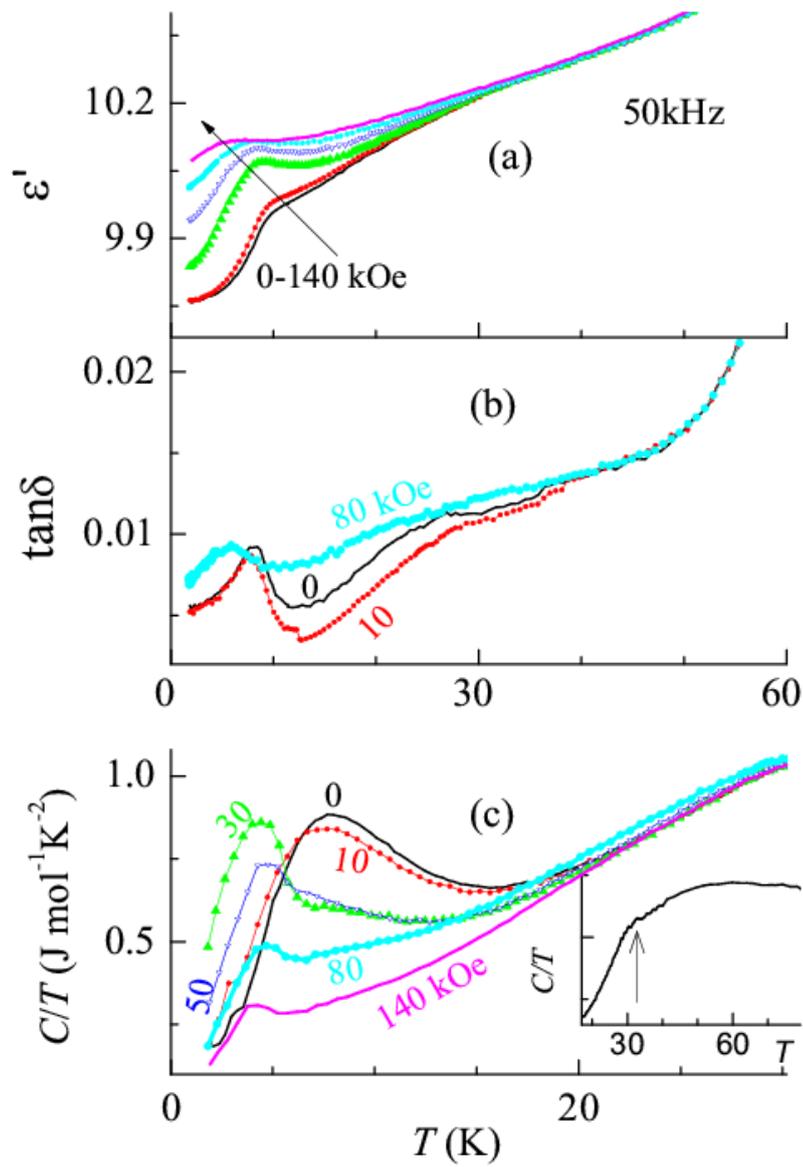

Figure 1:

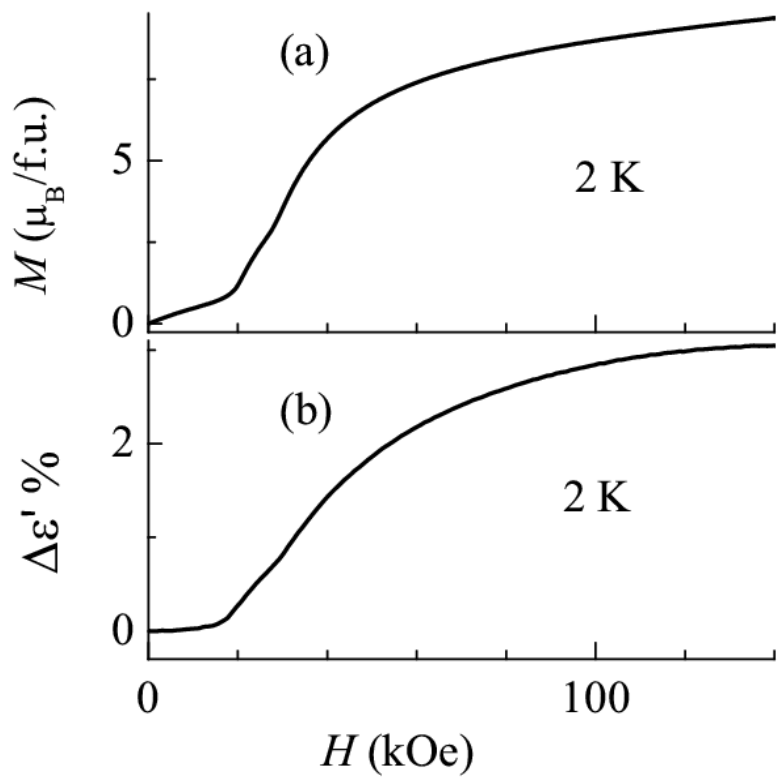

Figure 2:

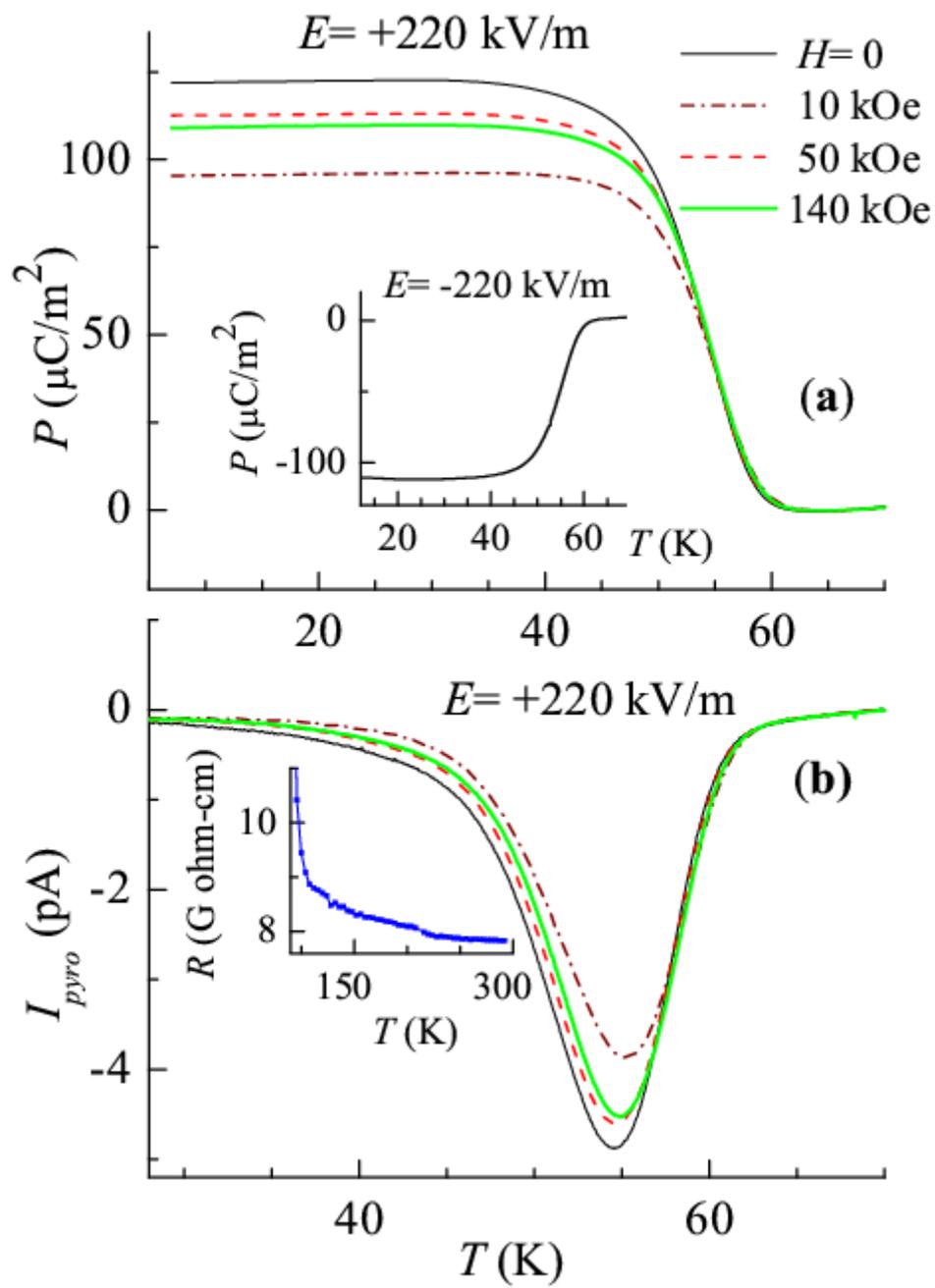

**Figure 3**

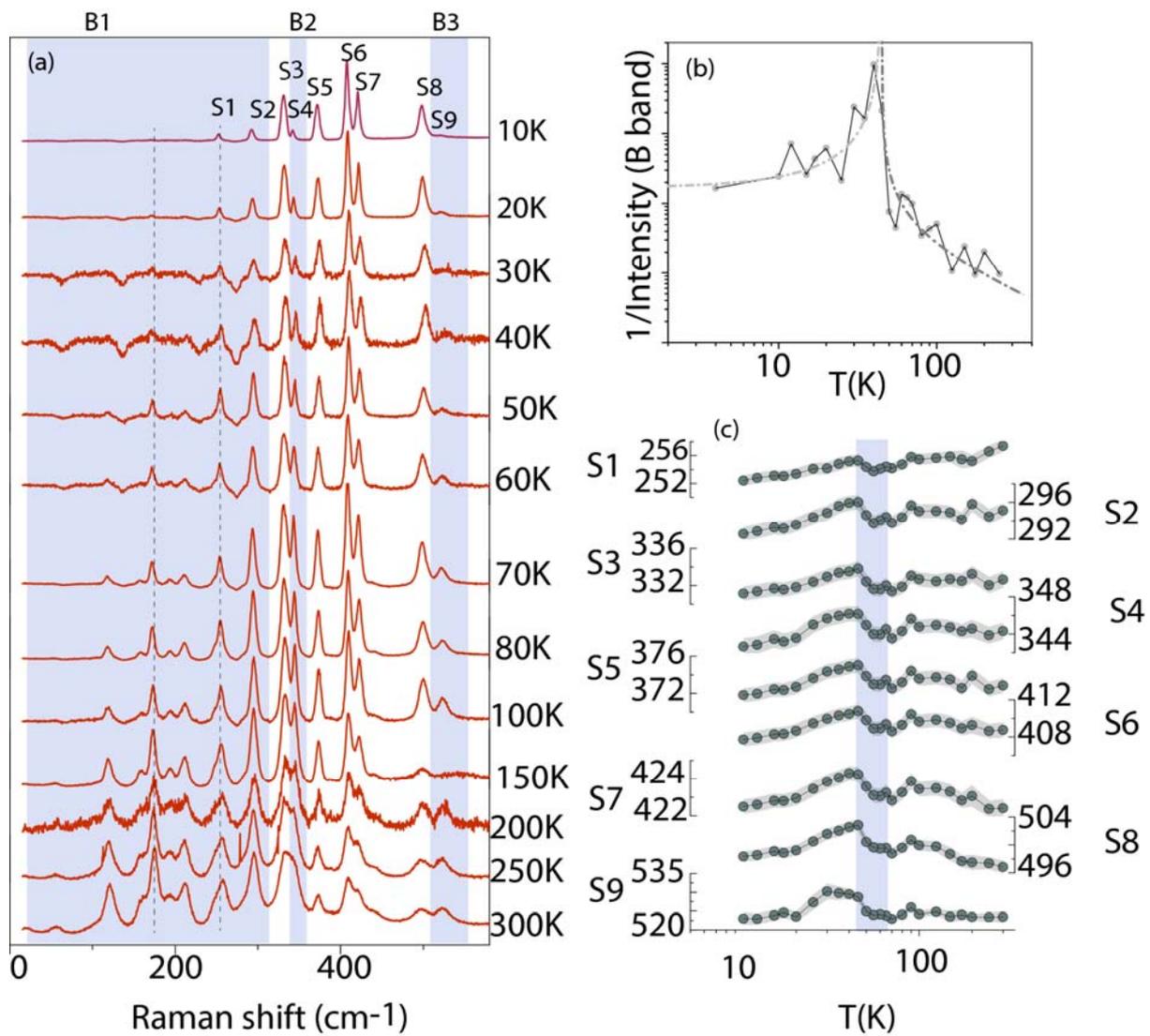

Figure 4

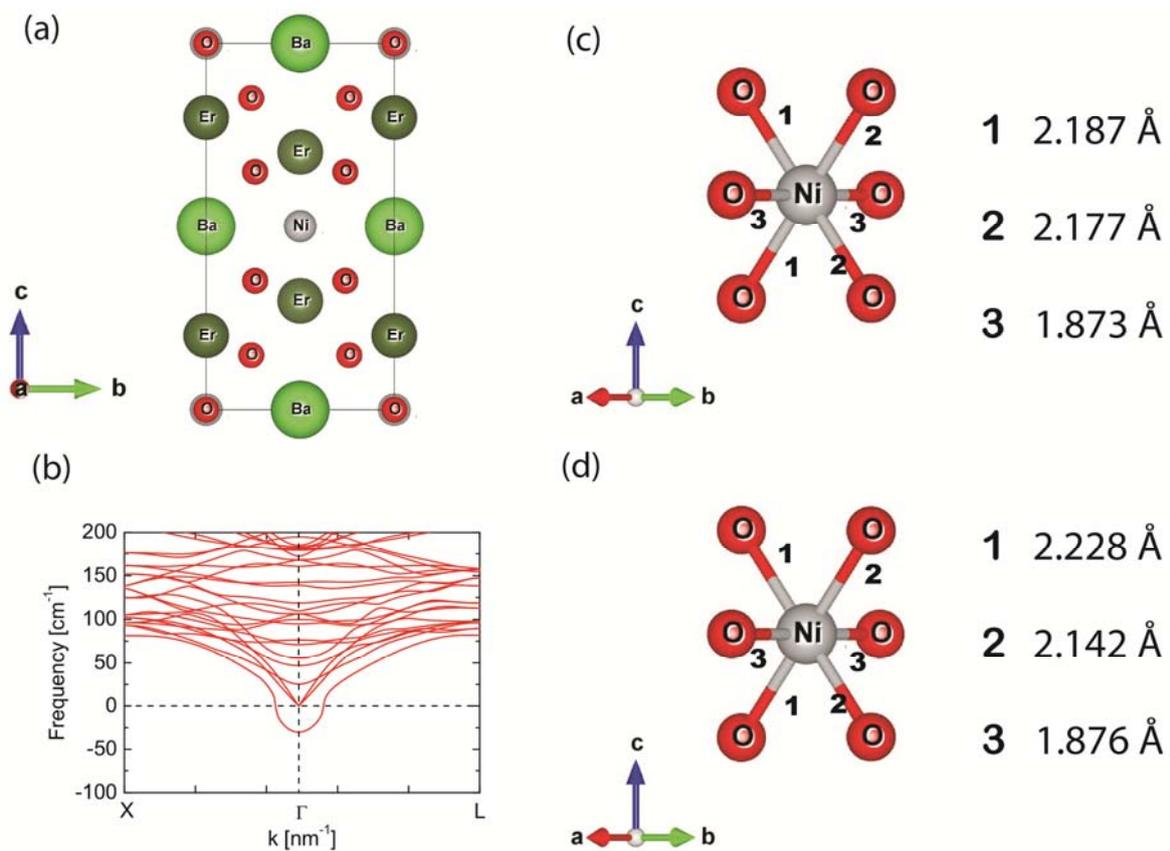

Figure 5

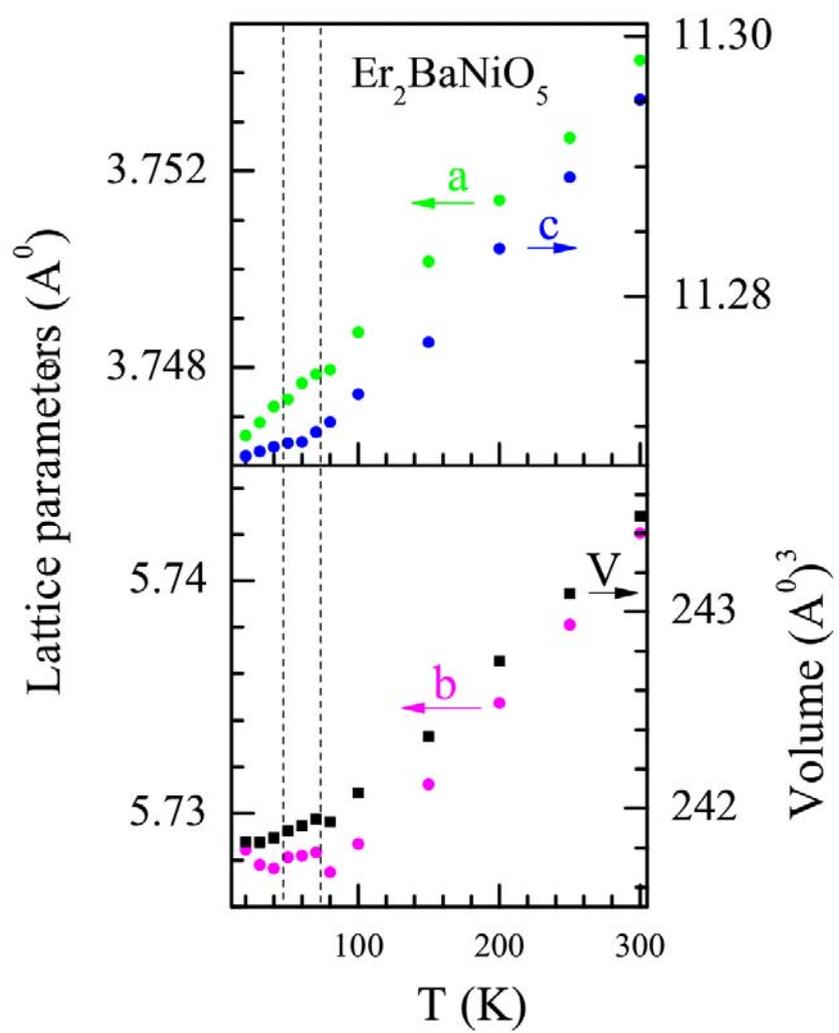

Figure 6

**Supplementary information**

**Displacive-type ferroelectricity from magnetic correlations within spin-chain**

Tathamay Basu,[1] V. V. Ravi Kishore,[2] Smita Gohil,[1] Kiran Singh,[1,3] N. Mohapotra,[4] S. Bhattacharjee,[1] Babu Gonde[1], N.P. Lalla,[3] Priya Mahadevan,[2] Shankar Ghosh,[1] and E.V. Sampathkumaran[1]

[1]Tata Institute of Fundamental Research, Homi Bhabha Road, Colaba, Mumbai-400005, India

[2]S. N. Bose National Centre for Basic Sciences, Sector-III, Block-JD, Salt Lake, Kolkata - 700 098, India

[3]UGC-DAE Consortium for Scientific Research, University Campus, Khandwa Road, Indore - 452001, India

[4]School of Basic Sciences, Indian Institute of Technology Bhubaneshwar, Bhubaneshwar-751013, India

Correspondence and requests for materials should be addressed to E.V.S (sampath@mailhost.tifr.res.in)

Here, we present x-ray diffraction pattern at 30 K and show the Rietveld fitting as well for three space groups discussed in the article. The refined parameters are also in the listed in the table.

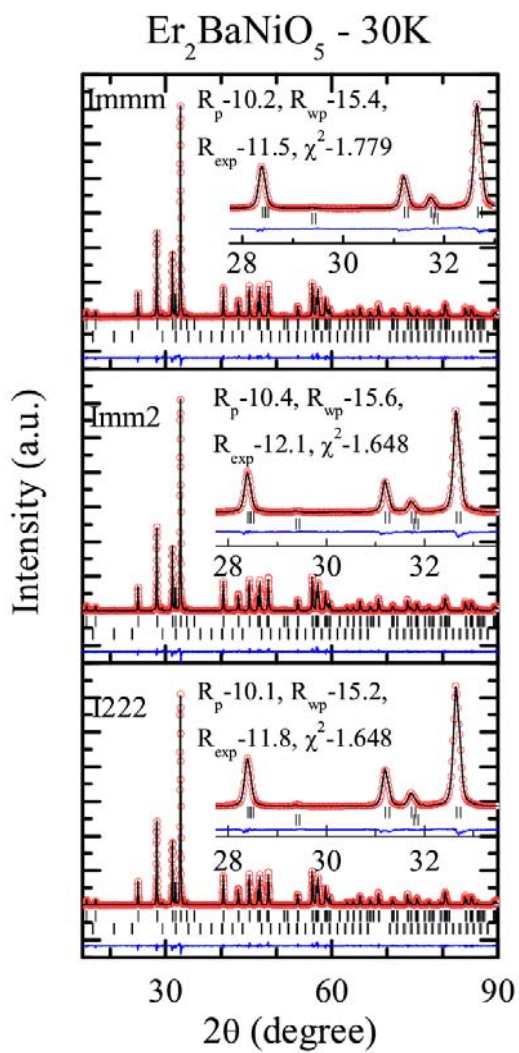

**Supplementary Figure S1:** The Rieveld fitted powder XRD pattern at 30 K. The fitting was carried out using FullProf program for three space groups. The parameters mentioned in the figure carry usual meaning. Observed (open red circle), calculated (continuous black line) and difference (continuous blue line) Upper vertical lines mark the Bragg positions of $Er_2BaNiO_5$; lower vertical lines represent Bragg positions of $Er_2O_3$ phase, assuming that it is present in small traces (<<2%). We also provide refined paramers below:

**Supplementary Table S1:** Rietveld refined structural parameters and statistical parameters

---

**Space group:** *Immm*

---

a= 3.7468(1) Å, b= 5.7277(1) Å, c= 11.26806(7) Å
$R_p$ = 10.4, $R_{wp}$ = 15.7, $R_e$ = 11.9 and $\chi^2$ = 1.759

---

|        | x   | y        | z        | B (Å$^2$) |
|--------|-----|----------|----------|-----------|
| Ba(2c) | 0.5 | 0.5      | 0        | 0.24(2)   |
| Er(4j) | 0.5 | 0        | 0.202(1) | 0.15(2)   |
| Ni(2a) | 0   | 0        | 0        | 0.29(4)   |
| O(8l)  | 0   | 0.240(2) | 0.148(3) | 0.1(3)    |
| O(2b)  | 0.5 | 0        | 0        | 0.1(4)    |

---

**Space group:** *Imm2*

---

a= 3.7466(1) Å, b= 5.7279(1) Å, c= 11.26816(7) Å
$R_p$ = 10.2, $R_{wp}$ = 15.2, $R_e$ = 11.8 and $\chi^2$ = 1.664

---

|         | x   | y        | z        | B (Å$^2$) |
|---------|-----|----------|----------|-----------|
| Ba(2a)  | 0.5 | 0.5      | 0        | 0.25(2)   |
| Er1(2b) | 0.5 | 0        | 0.202(1) | 0.16(4)   |
| Er2(2b) | 0.5 | 0        | 0.798(3) | 0.15(2)   |
| Ni(2a)  | 0   | 0        | 0.002(2) | 0.24(8)   |
| O11(4d) | 0   | 0.238(2) | 0.148(3) | 0.1(3)    |
| O12(4d) | 0   | 0.761(2) | 0.852(4) | 0.1(2)    |
| O2(2b)  | 0.5 | 0        | 0.003(1) | 0.1(1)    |

---

**Space group:** I222

---

a= 3.7468(4) Å, b=5.7278(4) Å, c=11.26811(7) Å
$R_p$ =10.1, $R_{wp}$ =15.2, $R_e$ =11.8 and $\chi^2$ = 1.648

|   | x | y | z | B (Å$^2$) |

------------------------------------------------------------------

| | | | | |
|---|---|---|---|---|
| Ba(2c) | 0.5 | 0.5 | 0 | 0.21(2) |
| Er(4j) | 0.5 | 0 | 0.202(1) | 0.16(2) |
| Ni(2a) | 0 | 0 | 0 | 0.31(4) |
| O(8k) | 0.000(2) | 0.240(2) | 0.148(3) | 0.1(7) |
| O(2b) | 0.5 | 0 | 0 | 0.1(7) |